# Tuning Carrier Type and Density in Highly Conductive and Infrared-Transparent (Bi$_{1-x}$Sb$_x$)$_2$Te$_3$ films

Xiangren Zeng[1,2,#], Shenjin Zhang[3,#], Zhiheng Li[1,2], Weiyue Ma[1,2], Renjie Xie[2], Yanwei Cao[2], Fengguang Liu[2], Fengfeng Zhang[3], Haichao Zhao[4,\*], and Xiong Yao[2,\*]

[1]School of Materials Science and Chemical Engineering, Ningbo University, Ningbo 315211, China

[2]Ningbo Institute of Materials Technology and Engineering, Chinese Academy of Sciences, Ningbo 315201, China

[3]Key Laboratory of Functional Crystals and Laser Technology, Technical Institute of Physics and Chemistry (TIPC), Chinese Academy of Sciences, Beijing 100190, China

[4]National Engineering Research Center for Remanufacturing, Army Arms University of PLA, Beijing 100072, China

Email: yaoxiong@nimte.ac.cn; zhchebei@sina.com

[#] These authors contribute equally to this work

ABSTRACT

Infrared transparent conductors have long been sought due to their broad optoelectronic applications in the infrared wavelength range. However, the search for ideal materials has been limited by the inherent trade-off between electrical conductance and optical transmittance. Band engineering offers an effective approach to modulate carrier type and density, enabling concurrent tuning of both conductance and transmittance. In this work, we present a band engineering strategy that enables effective tuning of both infrared transmittance and electrical conductance in topological insulator $(Bi_{1-x}Sb_x)_2Te_3$, bridging the gap and paving the way for applying topological insulators to infrared photoelectric devices. More importantly, with the combination of high carrier mobility and a large optical dielectric constant as suggested by previous report, $Sb_2Te_3$ achieves a high electrical conductance (∼1000 S/cm) and outstanding infrared transmittance (92.3%) in the wavelength range of 8~13 μm, demonstrating strong potential as an infrared transparent conductor. Our findings reveal that concurrent enhancement of both carrier mobility and optical dielectric constant is key to overcoming the conductance−transmittance trade-off. This work provides valuable insight for the exploration of high-performance infrared transparent conducting materials.

INTRODUCTION

Infrared transparent conductors are special optoelectronic materials that demonstrate high transmittance and excellent electrical conductivity in the infrared wavelength range[1–3], which are key components of various infrared optoelectronic applications, such as infrared thermal imaging, gas detection, safety detection, and infrared transparent electronics[4–11]. Currently, research on popular transparent conductors such as indium tin oxide (ITO) is limited to ultraviolet, visible, near-infrared, and mid-infrared wavelengths, while materials with both excellent conductivity and transmittance in the far infrared band are rare[12–15]. The high carrier concentration of ITO results in intense free carrier absorption, which reduces its transmittance in the far infrared wavelength range. The transparent wavelength range is limited by the plasma absorption edge ($\lambda_p$). Equations 1 and 2 clearly describe the trade-off dilemma between transparency and conductivity: $\lambda_p$ is inversely proportional to carrier density ($n$) and directly proportional to effective mass ($m^*$); while the conductivity ($\sigma$) is proportional to $n$ and inversely proportional to $m^*$. It is important to note that the relaxation time ($\tau$) and optical dielectric constant ($\varepsilon_{opt}$) are not entirely independent of carrier concentration ($n$) and effective mass ($m^*$). As a result, the key to solving this dilemma and achieving the desired infrared transparent conductor is to improve the dielectric constant in the optical frequency range ($\varepsilon_{opt}$) and the relaxation time ($\tau$), while keeping $n$ and $m^*$ almost unaffected[16,17].

$$\lambda_p = \frac{2\pi C_0}{\sqrt{m^*\varepsilon_0\varepsilon_{opt}/ne^2}} \propto \frac{m^*\varepsilon_{opt}}{n} \tag{1}$$

$$\sigma = \frac{ne^2\tau}{m^*} \propto \frac{n\tau}{m^*} \qquad (2)$$

Topological insulators, which are characterized by gapped bulk states and gapless surface states that are protected by time-reversal symmetry, have attracted intense research interest in the past decade, now offering a promising material platform to resolve the above dilemma for infrared-transparent conductors. Typical 3D topological insulators such as $Bi_2Se_3$, $Bi_2Te_3$, and $Sb_2Te_3$ are heavy-metal chalcogenides exhibiting a van der Waals structure comprised of a quintuple layer configuration[18–20]. The topological surface states in these compounds guarantee a high mobility, which means a long carrier relaxation time[21–23]. At the same time, these compounds satisfy the three major principles of low ionization energy, low hybridization, and low saturation, leading to a high $\varepsilon_{opt}$[16]. So the two requirements of high $\tau$ and $\varepsilon_{opt}$ for infrared transparent conductors are perfectly fulfilled in these compounds, especially in $Bi_2Te_3$ and $Sb_2Te_3$. More importantly, the carrier density and mobility can be precisely controlled by a band engineering strategy, i.e., tuning the chemical doping in the ternary alloy $(Bi_{1-x}Sb_x)_2Te_3$[24–28]. By adjusting the Fermi level via band engineering, the interference of free carriers on infrared absorption can be manipulated, enabling tunable infrared transmittance. We employed the atomic layer-by-layer molecular beam epitaxy (MBE) technique to grow high-quality $(Bi_{1-x}Sb_x)_2Te_3$ thin films. This technique offers a low and controllable growth rate, precisely controlled doping level at atomic-scale, and low growth temperatures[29–31], leading to high mobility (high $\tau$) epitaxial films. Our results show that the carrier density and mobility can be effectively adjusted through bandgap engineering in $(Bi_{1-x}Sb_x)_2Te_3$ thin films, with the infrared transmittance improved to above 90% and the conductivity exceeding 1000 S/cm.

RESULTS AND DISCUSSION

In this study, $(Bi_{1-x}Sb_x)_2Te_3$ films were prepared on double-side polished $BaF_2$ (111) substrates by molecular beam epitaxy. The $BaF_2$ (111) substrate was chosen due to its high transmittance of nearly 92.7% in the 8~13 μm far-infrared wavelength range and a small lattice mismatch with $Bi_2Te_3$. Figure 1(a) shows the crystal structure of $Bi_2Te_3$ ($Sb_2Te_3$). Both $Sb_2Te_3$ and $Bi_2Te_3$ share the same crystal structure ($R\bar{3}m$) with a quintuple layer configuration of Te-Bi-Te-Bi-Te, in which the Sb atom can easily replace the Bi atom to form the compound $(Bi_{1-x}Sb_x)_2Te_3$ with an arbitrary x ratio. In $Bi_2Te_3$, the conductance is n-type, attributed to electron carriers induced by Te vacancies. In contrast, the p-type conductance in $Sb_2Te_3$ is dominated by Sb-Te antisite defects. The alloying in $(Bi_{1-x}Sb_x)_2Te_3$ enables an effective tuning of carrier type and density over a wide range. The crystallinity of the $(Bi_{1-x}Sb_x)_2Te_3/BaF_2$ thin films was characterized by X-ray diffraction (XRD) scanning, as shown in Figure 1(b). All $(Bi_{1-x}Sb_x)_2Te_3$ (00 3l) diffraction peaks are clearly visible, and the XRD patterns are labeled by Sb doping level $x$ from 0 to 1 from bottom to top. The Sb doping concentration $x$ for all samples was determined by X-ray photoelectron spectroscopy (XPS) measurements (Figure S1 and Table S1). At low Sb doping levels, the films exhibit similar diffraction peaks to $Bi_2Te_3$, mainly because the lattice constants of $Bi_2Te_3$ and $Sb_2Te_3$ are very close, resulting in no significant shift in the diffraction peaks. However, as the amount of antimony doping increases, a weak (009) diffraction peak occurs when Sb content reaches $x$=0.45, and this peak becomes clear when antimony completely replaces all the bismuth. The (009) peak is a forbidden reflection for $Bi_2Te_3$ due to the XRD extinction rule. However, this peak is allowed for $Sb_2Te_3$ because the substitution of Sb for Bi changes the structure factor, thereby breaking the extinction law. We further characterized the surface topography of a 20nm $Bi_2Te_3$/$BaF_2$ using atomic force microscopy (AFM), as shown in Figure 1(c). The surface exhibits typical terraces in a regular triangular shape, indicating high quality of the sample[32].

To evaluate the conductance performance of these $(Bi_{1-x}Sb_x)_2Te_3$ films, we conducted electrical transport measurements using the Physical Property Measurement System (PPMS). Figure 2(a) shows the temperature-dependent conductivity of these samples. For $Bi_2Te_3$ ($x=0$) and $Sb_2Te_3$ ($x=1$), the conductivity shows a metallic behavior with decreasing temperature, owing to the carriers induced by Te vacancies or antisites in the bulk states. It is worth mentioning that the conductivity of $Sb_2Te_3$ reaches almost 1000 S/cm at room temperature and above 2000 S/cm at 2 K, exceeding the application requirements for infrared transparent conductors. With increasing Sb doping level in $Bi_2Te_3$, the conductivity initially decreases due to carrier compensation between n-type $Bi_2Te_3$ and p-type $Sb_2Te_3$. It reaches a minimum value at $x=0.62$, a signature of the n-p charge neutral point (CNP), and increases again at $x>0.62$ owing to increased p-type carriers[24,26,28]. Figure 2(b) shows a plot of the field-dependent conductivity of the films measured at 2 K, in which the conductivity is multiplied by different coefficients to make the features clear. All the sample exhibits a clear cusp near the zero magnetic field, a signature of the weak antilocalization (WAL) effect, which is related to the emergence of topological surface states[33,34]. Figures 2(c) and 2(d) show the Hall resistance measured at 2 K and 300 K, respectively. These results clearly show that with the increase of Sb doping, the carrier type changes from n-type to p-type. All the Hall resistance exhibit a distinctly linear behavior, probably indicating the conductance of these films are dominated by single type of carrier, even though this possibility requires further verification by high magnetic field measurements. We can clearly see a sign change from n-type with $x$=0.62 to p-type with $x$=0.81 in Hall resistance, confirming that $x$=0.62 is the CNP. Our observations of conductivity and Hall resistance in Figure 2 are consistent with the previous reports[24,28]. We successfully attained highly tunable carrier density and conductance in $(Bi_{1-x}Sb_x)_2Te_3/BaF_2$ films, and realized conductivity as high as 2000S/cm in $Sb_2Te_3$.

We extracted the carrier density and mobility of all the $(Bi_{1-x}Sb_x)_2Te_3$ films from Figure 2 and plotted them in Figures 3(a) and (b). At 2 K, the carrier density reaches a minimum value of $|n_{3D}|=1.98 \times 10^{18}$ cm$^{-3}$ at $x$=0.62, while at 300 K, the minimum value is $|n_{3D}|=5.39 \times 10^{18}$ cm$^{-3}$ at $x$=0.45, showing a similar trend for both temperatures. In previous reports about $(Bi_{1-x}Sb_x)_2Te_3$ films, the mobility typically shows either a minimum or maximum at the CNP[24,28], but this is not the case in our $(Bi_{1-x}Sb_x)_2Te_3/BaF_2$ films. As shown in Figures 3(a) and 3(b), the mobility generally increases from $Bi_2Te_3$ to $Sb_2Te_3$, except for an abrupt change at $x = 0.45$. This abnormal mobility trend is likely due to the mixed contribution from both p-type and n-type carriers, while the exact origin still requires further investigation. Nevertheless, our $Sb_2Te_3$ thin films exhibit high mobilities of 623.9 cm²/V·s at 2 K and 388.5 cm²/V·s at 300 K, outperforming previously reported infrared-transparent conducting materials such as Bi$_2$Se$_{2.4}$ (113.8 cm²/V·s) and Bi$_2$Se$_{1.8}$Te$_{0.7}$ (86.8 cm²/V·s)[16]. This high mobility suggests a long relaxation time $\tau$, which explains the outstanding conductance of Sb$_2$Te$_3$ in Figure 2(a). Then we checked the band structure of a (Bi$_{0.38}$Sb$_{0.62}$)$_2$Te$_3$/TiN film using a home-built angle-resolved photoemission spectroscopy (ARPES) system, as shown in Figure 3(c). TiN was selected as the substrate due to the conducting requirements for ARPES measurements. The ARPES spectra clearly show a V-shaped Dirac surface state, confirming the non-trivial topology of the (Bi$_{0.38}$Sb$_{0.62}$)$_2$Te$_3$ layer. The Dirac point lies as close as 0.09 eV below the Fermi level, implying close proximity to the CNP, which is consistent with our transport results.

We further performed Fourier transform infrared spectroscopy (FTIR) measurement on the prepared thin film samples with the same thickness of (20±2) nm to test the infrared transmittance, and the results are shown in Figure 4(a). Here, we focus on the infrared transmittance in the far-infrared wavelength range (8~13 μm). utilizing BaF$_2$ as substrates with an average transmittance of 92.7% in this range. The transmittance of all the (Bi$_{1-x}$Sb$_x$)$_2$Te$_3$ thin films is calculated by Equation $T_{film} = \frac{T_{total}}{T_{substrate}}$ to eliminate the influence of the BaF$_2$ substrate on the transmittance. Combining Figures 3b and 4b, it can be clearly seen that the infrared transmittance can be tuned by carrier density, but the change is considerably smaller than that of the conductance. For example, in the samples with $x$ = 0.45, 0.62, and 0.81, the conductance is strongly suppressed by band engineering of the carrier density, whereas the infrared transmittance exhibits only a moderate improvement compared to Bi$_2$Te$_3$. Notably, Sb$_2$Te$_3$ exhibits the highest infrared transmittance among all the (Bi$_{1-x}$Sb$_x$)$_2$Te$_3$ samples, even though its carrier concentration is not the lowest. Therefore, the above observations suggest that carrier density is not the only factor affecting transmittance, highlighting the significant role of $\varepsilon_{opt}$, as shown in Equation 1. According to a previous report, $Sb_2Te_3$ has a substantially higher $\varepsilon_{opt}$ than $Bi_2Te_3$[16], which explains its unusually high infrared transmittance. The average relative transmittance is 92.3% for $Sb_2Te_3$ in the entire 8~13 μm far-infrared wavelength range, which is higher than that of the previously reported InHfO[35]. In addition, we also compared the transmittance of MnTe films, $Bi_4Te_3$ films, and $(Bi_{1-x}Sb_x)_2Te_3$ films, as shown in Figure 4(a). MnTe film exhibits higher transmittance than all $(Bi_{1-x}Sb_x)_2Te_3$ films, but suffers from extremely high electrical resistance and poor conductivity. In contrast, $Bi_4Te_3$ shows the lowest transmittance despite having excellent conductance. By comparison, the $(Bi_{1-x}Sb_x)_2Te_3$ films not only maintain good electrical conductivity but also exhibit excellent infrared transmittance. As summarized in Figure 4(b), these films demonstrate a clear trade-off between transmittance and conductance for Sb doping level from $x =$ 0.25 to 0.81, where an increase in one value generally corresponds to a decrease in the other. The only exceptions to this trend are $Bi_2Te_3$ and $Sb_2Te_3$. Overall, the band engineering strategy via Sb doping is more effective in tuning electrical conductance than infrared transmittance. When the influence of $\varepsilon_{opt}$ is taken into account, $Sb_2Te_3$ emerges as a promising candidate for infrared transparent conductor, combining both high conductance (around 1000 S/cm) and high infrared transmittance (92.3% on average for 8~13 μm wavelength).

To comprehensively evaluate the infrared transparent conducting properties of the $(Bi_{1-x}Sb_x)_2Te_3$ films, we employed the transparent conductivity figure of merit (FOM), defined as FOM = $-1/(R_\square \cdot \ln T)$, where T is the transmittance (0 < T < 1) and $R_\square$ is the sheet resistance. The FOM values were calculated in the wavelength range of 8~13 μm, as shown in Figure 4(c). $Sb_2Te_3$ exhibits significantly superior FOM of (2.35 ×$10^{-2}$) compared to all other samples, outperforming values reported in materials such as $Bi_2Se_{2.4}$(5.35×$10^{-3}$), $Bi_2Se_{1.8}Te_{0.7}$(5.41×$10^{-3}$), $PbSe_{1.6}$(4.67×$10^{-3}$), and $Bi_2Te_{2.4}$(9.42×$10^{-3}$)[16]. This enhanced performance can be attributed to the combination of high mobility, which results in excellent conductance, and a high $\varepsilon_{opt}$, which contributes to strong transmittance. $Bi_2Te_3$ demonstrates the second-highest FOM, outperforming all the other doped $(Bi_{1-x}Sb_x)_2Te_3$ films. In contrast, the sample with $x$ = 0.62, which lies near the CNP, shows the lowest FOM, likely due to a more pronounced suppression of conductance than the improvement in transmittance by carrier density tuning.

CONCLUSIONS

In this work, we demonstrate that band engineering is an effective strategy for tailoring the infrared optoelectronic properties of $(Bi_{1-x}Sb_x)_2Te_3$ thin films, although its influence on electrical conductance is more substantial than on infrared transmittance. With the combination of a high carrier mobility (or relaxation time) and a high optical dielectric constant $\varepsilon_{opt}$ as suggested by previous report[16], $Sb_2Te_3$ simultaneously achieves excellent electrical conductance and high infrared transmittance, showcasing strong potential as an infrared-transparent conductor. Our findings reveal that enhancing both mobility and $\varepsilon_{opt}$ is essential to overcome the conventional trade-off between conductance and transmittance. Furthermore, given that $Sb_2Te_3$ exhibits superior conductance of ~1000 S/cm while MnTe offers high infrared transmittance of 97.08% in the 8~13 μm range, we propose $MnSb_2Te_4$ as an ideal infrared-transparent conducting material and warrants further investigation, considering this recently emerged magnetic topological insulator is a natural superlattice composed of these two materials.

METHODS

Preparation of the $(Bi_{1-x}Sb_x)_2Te_3/BaF_2$ thin films: We grow all the samples on double-side polished 5 mm × 5 mm $BaF_2$ (111) substrates in a custom-made ultra-high vacuum MBE system with base pressure lower than $5 \times 10^{-10}$ Torr. Prior to sample growth, the substrate was ultrasonically cleaned with acetone and anhydrous ethanol for 10 minutes, dried with a nitrogen gun, and then placed into the MBE chamber and annealed at 450 °C for 15 minutes to clean the surface. Pure elemental Bi (99.997%), Sb (99.999%), and Te (99.9999%) sources were evaporated by Knudsen cells. To minimize Te vacancies, growth was performed with a (Bi, Sb) to Te flux ratio of 1:10. All source fluxes were in situ calibrated using a quartz crystal microbalance. We employed a two-step growth method to grow the films: first, depositing four quintuple layers (QLs) of $(Bi_{1-x}Sb_x)_2Te_3$ at 150 °C as the seed layer, and then growing the rest at 240 °C with a fixed total thickness of (20±2) nm.

Crystal structure and surface analysis: The XRD measurements were performed on a high-resolution X-ray diffraction (Bruker D8 Discovery HRXRD) with Cu Kα irradiation (λ=1.5406 Å). Symmetric $\theta$-$2\theta$ scanning and X-ray reflectance (XRR) analysis were performed under 40 kV/40 mA operating conditions. Surface topography was characterized by a Bruker Dimension Icon atomic force microscope (Digital Instruments, USA) with Nanoscope V controller.

Transport Measurement: Electrical transport properties of $(Bi_{1-x}Sb_x)_2Te_3$ thin films were measured using a Physical Properties Measurement System (PPMS). Measurements of the longitudinal resistivity ($R_{xx}$) and Hall resistance ($R_{xy}$) were performed simultaneously using the van der Pauw configuration, with four indium electrodes mechanically contacted at the corners of the sample.

Angle-Resolved Photoemission Spectroscopy Measurement: The band structure of the $(Bi_{0.38}Sb_{0.62})_2Te_3$/TiN heterostructure was measured by a homemade ARPES system with a 177 nm deep-UV laser source ($h_v$ =6.997 eV). Measurements were conducted under ultrahigh vacuum with a total energy resolution better than 1 meV at 7.3 K. The Fermi level was determined by a polycrystalline gold standard sample in electrical contact with the sample.

Fourier Transform Infrared Transmission Measurement: The infrared transmittance of the thin films was measured by NICOLET 6700 FTIR spectrometer (Thermo, USA). Measurements were conducted in transmission mode, where infrared light was perpendicularly incident on the thin film surface, and the transmitted light was collected to record the transmittance spectrum. The spectral range was set from 4000 to 400 cm$^{-1}$. All measurements were performed at room temperature under ambient conditions.

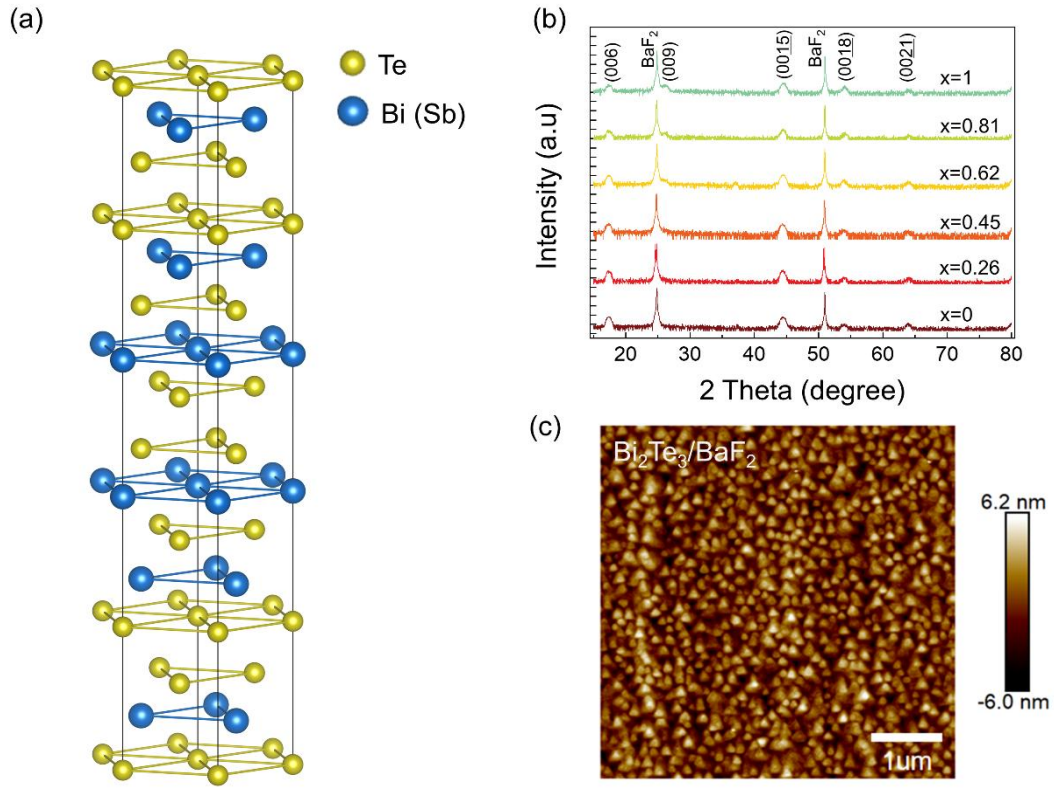

Figure 1. (a) Crystal structure of $Bi_2Te_3(Sb_2Te_3)$. (b) XRD profiles of $(Bi_{1-x}Sb_x)_2Te_3$ with $x$=0, 0.26, 0.45, 0.62, 0.81, and 1. The thickness of the $(Bi_{1-x}Sb_x)_2Te_3$ thin films in Figure 1 is (20±2) nm. (c) Atomic force microscopy (AFM) topography of a typical $Bi_2Te_3$ grown on $BaF_2$(111) with a thickness of approximately 20 nm.

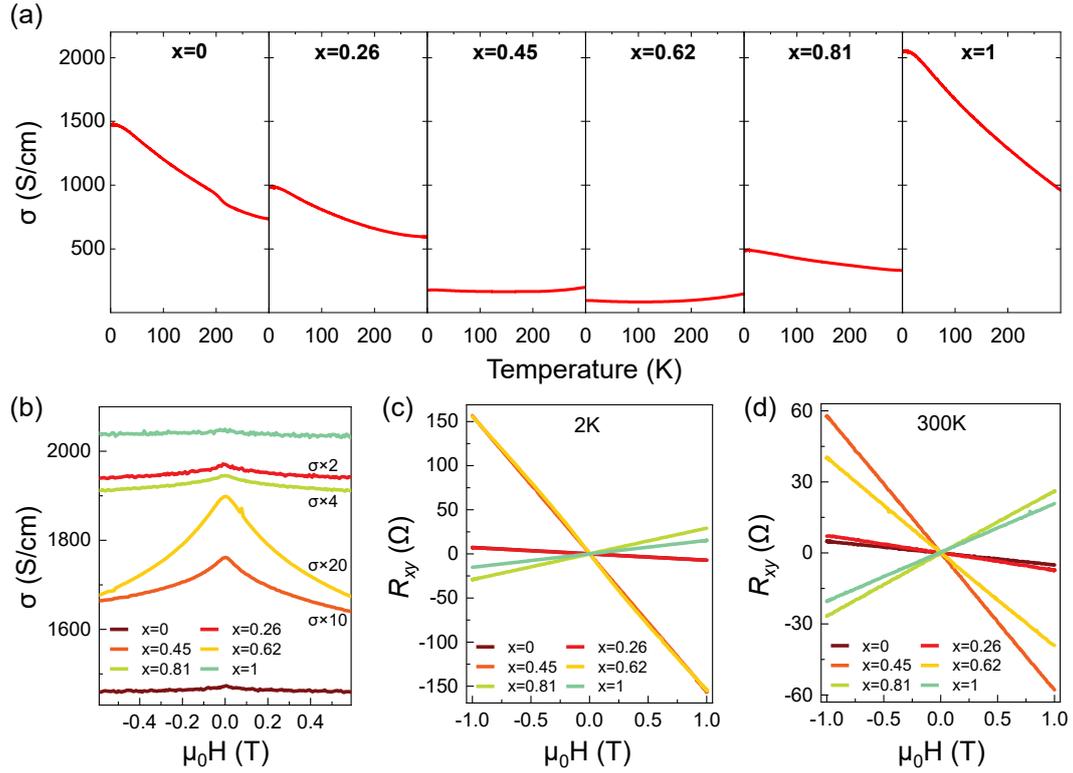

Figure 2. Electrical transport measurements of $(Bi_{1-x}Sb_x)_2Te_3$ thin films grown on $BaF_2(111)$ ($x$=0, 0.26, 0.45, 0.62, 0.81, 1). (a) Temperature-dependent conductivity measured in the range of 300 K to 2 K. (b) Magnetic Field-dependent conductivity measured at 2 K and up to 0.6 T. (c) Magnetic Field-dependent Hall resistance $R_{xy}$ measured at 2 K and up to 1 T. (d) Magnetic Field-dependent Hall resistance $R_{xy}$ measured at 300 K and up to 1 T.

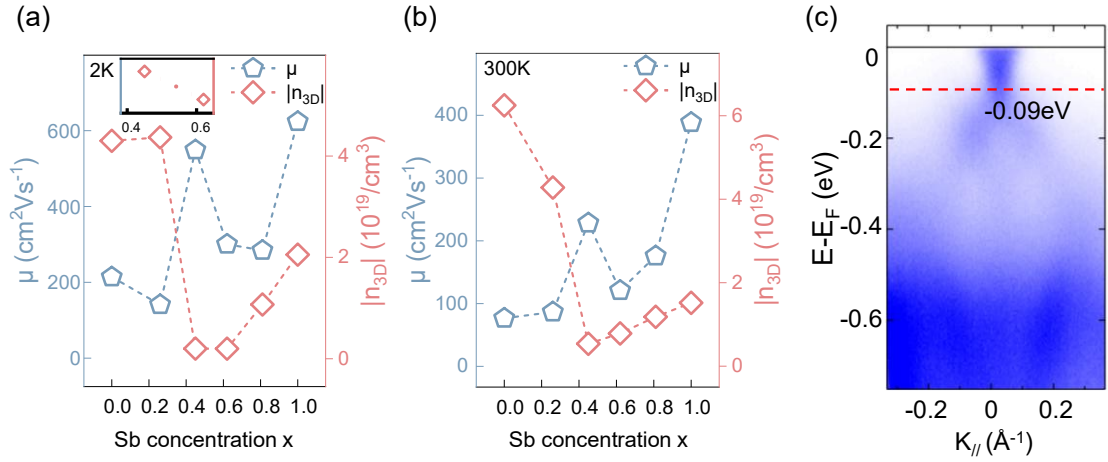

Figure 3. (a, b) Mobility and carrier density of $(Bi_{1-x}Sb_x)_2Te_3$ films extracted from conductivity and Hall resistance at (a) 2 K and (b) 300 K, respectively. (c) The ARPES spectra of a 20 nm thick $(Bi_{0.38}Sb_{0.62})_2Te_3$ film grown on TiN. Data were collected at 7.3 K.

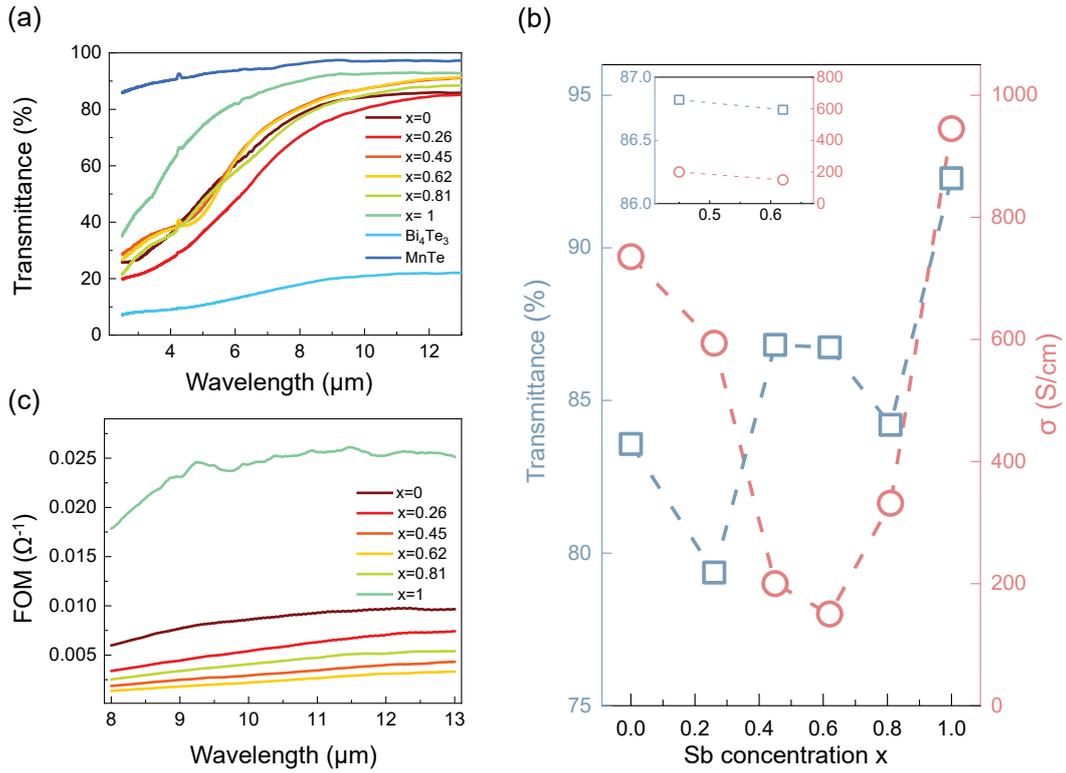

Figure 4. (a) The infrared transmission spectra of $(Bi_{1-x}Sb_x)_2Te_3$ films with varied Sb doping level $x$, MnTe film, and $Bi_4Te_3$ film. (b) The summary of infrared transmittance and conductivity for all the $(Bi_{1-x}Sb_x)_2Te_3$ samples. The transmittance is the averaged value in the entire 8~13 μm far-infrared wavelength range. (c) The transparent conductivity figure of merit (FOM) of all the $(Bi_{1-x}Sb_x)_2Te_3$ films.

Supporting information:

The XPS high-resolution spectra of the (Bi$_{1-x}$Sb$_x$)$_2$Te$_3$ films; the AFM images of the (Bi$_{1-x}$Sb$_x$)$_2$Te$_3$ films

AUTHOR INFORMATION

Corresponding Authors

*E-mail: yaoxiong@nimte.ac.cn;

zhchebei@sina.com

Author Contributions

# X.Z. and S.Z. contribute equally to this work. X.Y. and H.Z. conceived the experiments. X.Z., W.M., R.X. and X.Y. grew the (Bi$_{1-x}$Sb$_x$)$_2$Te$_3$ and MnTe thin films. Z.L. grew the Bi$_4$Te$_3$ films. X.Z., and X.Y. performed the electric transport and infrared transmittance measurements. X.Z., S.Z., F.Z. and R.X. performed and analyzed the ARPES measurements. X.Y., H.Z., Y.C., F.L. and X.Z. analyzed the results and wrote the manuscript with contributions from all authors.

Notes

The authors declare no competing financial interest.

ACKNOWLEDGMENT

The work is supported by the National Natural Science Foundation of China (Grant Nos. 12304541), the National Key Research and Development Program of China (Grant No. 2024YFF0508500), the Hundred Talents Program of Chinese Academy of Sciences, and the Ningbo Science and Technology Bureau (Grant Nos. 2023J047).

**For TOC only**

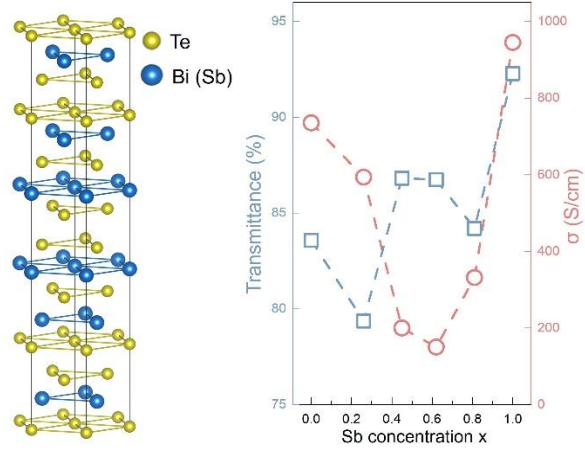